\documentclass{article}
\usepackage{spconf,amsmath,graphicx}
\usepackage{times}
\usepackage{epsfig}
\usepackage{graphicx}
\usepackage{subcaption}
\usepackage[dvipsnames]{xcolor}
\usepackage{amsmath}
\usepackage{amssymb}
 
\usepackage{siunitx}
\usepackage{float}
\usepackage{subcaption}

\let\vec\mathbf
\newcommand{\norm}[1]{\Vert #1 \Vert}
\newcommand{\T}{^\mathsf{T}}
\newcommand{\R}{\mathbb{R}}

\title{Multiple Offsets Multilateration: a new paradigm for sensor network calibration with unsynchronized reference nodes}
\name{Luca Ferranti $^{a,c}$, Kalle Åström $^b$, Magnus Oskarsson $^b$, Jani Boutellier $^a$, Juho Kannala $^c$}
\address{$^a$University of Vaasa, Vaasa, Finland, $^b$Lund University, Lund, Sweden\\ $^c$Aalto University, Espoo, Finland}
\begin{document}
%\ninept
%
\maketitle
\begin{abstract}
Positioning using wave signal measurements is used in several applications, such as GPS systems, structure from sound and Wifi based positioning. Mathematically, such problems require the computation of the positions of receivers and/or transmitters as well as time offsets if the devices are unsynchronized. In this paper, we expand the previous state-of-the-art on positioning formulations by introducing Multiple Offsets Multilateration (MOM), a new mathematical framework to compute the receivers positions with pseudoranges from unsynchronized reference transmitters at known positions. This could be applied in several scenarios, for example structure from sound and positioning with LEO satellites. We mathematically describe MOM, determining how many receivers and transmitters are needed for the network to be solvable, a study on the number of possible distinct solutions is presented and stable solvers based on homotopy continuation are derived. The solvers are shown to be efficient and robust to noise both for synthetic and real audio data.
\end{abstract}

\section{Introduction}
\label{ch:intro}
Given a network of receivers measuring a wave signal emitted by some transmitters, the \textit{network calibration} problem is concerned with determining the positions of receivers and/or transmitters (together more generally referred as \textit{nodes}) \cite{taylor2006simultaneous}. This is a general positioning scenario which can arise in several applications. In GPS systems, one aims at computing the receiver position, knowing the satellite transmitter positions \cite{wells1987guide}. For indoor localization applications, one may try to locate the user position from measured wifi signals \cite{bell2010wifi}. In structure from sound applications \cite{thrun2005affine}, one aims at simultaneously determining microphones and loudspeakers.

Depending on what node positions are known, and whether or not the nodes are synchronized, several scenarios can arise. In the general trilateration problem \cite{thomas2005revisiting}, one computes the receiver position from measured distances from transmitters at known positions. For GPS positioning, a variant of trilateration where the receiver is also unsynchronized is used \cite{bancroft1985algebraic}. 
For the network self-calibration scenario, where both receivers and transmitters are at unknown positions, several formulations have been proposed, depending on the degree of synchronization. In the Time Of Arrival (TOA) formulation, all nodes are assumed to be synchronized and their positions are computed from measured pseudoranges \cite{kuang2013complete, burgess2015toa}. In Time Difference Of Arrival (TDOA) \cite{kuang2013stratified}, the transmitters are assumed to be unsynchronized and in Unsynchronized TDOA (UTDOA) all nodes are unsynchronized  \cite{burgess2012node}.

During the years, various numerical techniques have been explored to solve different positioning scenarios. In some situations, a closed form solution is possible \cite{bancroft1985algebraic, moses2003self}. Several iterative algorithms, based on minimizing some loss function, have also been widely explored \cite{biswas2004passive, priyantha2003anchor, rahman2017trilateration}. As the positioning problems described usually can be  formulated as systems of polynomial equations, more recently,  approaches based on algebraic geometry have also been proposed \cite{kuang2013complete, kuang2013stratified, ferranti2021sensor, ferranti2021homotopy}.

In this paper we introduce \textit{Multiple Offsets Multilateration (MOM)}, a new network calibration paradigm where we want to compute the positions of a set of synchronized receivers using a set of unsynchronized transmitters and measuring only the time at which the signal is received. This new framework can be thought of as a generalization of trilateration (now the transmitters are unsychronized) or equivalently as a special case of TDOA (now the transmitters are at known positions). We argue that this new positioning scenario could find plenty of applications. For example, Low Earth Orbit (LEO) satellites having potential for positioning have recently attracted more interest in research \cite{khalife2019receiver}. Opposed to traditional GNSS systems, LEO satellites have a less accurate clock \cite{hauschild2008real} and hence in addition to the receiver position also the satellite time offset needs to be estimated \cite{khalife2019receiver}. On the other hand, several models for LEO satellites orbit determination have been proposed \cite{vsvehla2003kinematic, bennett2013accurate} and hence the Multiple Offsets Multilateration technique presented in this paper could offer an appealing technique for LEO positioning.

The contribution of our work can be summarized as follows
\begin{itemize}
    \item We introduce \textit{Multiple Offsets Multilateration}, a new paradigm for positioning that can be used to compute receivers positions using the radio signal measured from unsynchronized transmitters at known position.
    \item We give a mathematical formulation of MOM and present a full characterization. That is, we determine how many nodes a MOM network must have at least to be solvable and using computational algebraic geometry determine rigorously the \textit{degree} of the problem, that is how many solutions it can have in total.
    \item We propose numerical recipes that allow to solve efficiently and robustly different MOM networks in different scenarios.
\end{itemize}

This paper is structured as follows. In Section \ref{ch:formulation} we give a mathematical formulation of MOM and derive a full characterization of the framework. In Section \ref{ch:method}, the proposed method to solve MOM problems is derived. Finally, we describe the numerical experiments in Section \ref{ch:results} and draw conclusions in Section \ref{ch:conclusions}.

% \section{Related Work}
% \label{ch:relatedWork}
% \input{tex/2relatedWork}

\section{Problem Formulation}
\label{ch:formulation}
In this section we describe the mathematical formulation of \textit{Multiple Offsets Multilateration} (MOM). Suppose we have $m$ receivers at positions $\vec{r}_1,\ldots,\vec{r}_m$ and $n$ transmitters at positions $\vec{s}_1,\ldots,\vec{s}_n$. We denote such a network shortly as $m$r/$n$s. Furthermore, suppose that the signal from the transmitter to the receiver travels with (known) constant speed $v$. Let us also assume that all receivers are synchronized. Then for each $i=1,\ldots,m$ and each $j=1,\ldots,n$ we can write 

\begin{equation}
    \norm{\vec{r}_i-\vec{s}_j} = v(t_{ij} - t_j),
\end{equation}
where $t_j$ is the time instant at which the signal was sent from the transmitter $j$ and $t_{ij}$ is the measured time instant at which the signal is received at receiver $i$. This equation can be rewritten as

\begin{equation}
     \norm{\vec{r}_i-\vec{s}_j}^2 = (f_{ij} - o_j)^2, \label{eq:range}
\end{equation}
where $f_{ij}$ is now the measured pseudorange between the $i$th receiver and $j$th transmitter and $o_j$ is the offset corresponding to the transmitter. Now we want to estimate the positions of the receivers $\vec{r}_i$ and the offsets of the transmitters $o_j$. Hence we will have $2m+n$ unknowns in 2D and $3m+n$ unknowns in 3D. From the measurements we have $mn$ equations of type \eqref{eq:range} in total.

To study the solvability of MOM we define the \textit{excess constraint} as

\begin{equation}
    c = mn - Km - n,
\end{equation}
where $K$ is the dimension of the space ($K=2$ or $K=3$). If $c < 0$, then the problem is \textit{underdetermined}, that is it has an infinite number of possible solutions and the network is not solvable. If $c > 0$, then the problem is \textit{overdetermined} and a solution, assuming it exists, will be unique. The case $c=0$ leads to a \textit{determined} configuration, that is it will have a finite, but not necessarily unique, number of solutions. These configurations are the smallest networks which are not underdetermined and are hence called \textit{minimal configurations}. Table \ref{tab:confs} shows the underdetermined, minimal and determined configurations for MOM in 2D and 3D.

\begin{table}[tbh]
    \centering
        \caption{MOM configurations for 2D (left) and 3D (right). u: underdetermined. M: minimal. *: overdetermined.}
    \label{tab:confs}
    \begin{tabular}{c||ccccc}
        $m\backslash n$&2&3&4&5\\\hline\hline
        2&u&u&M&*\\
        3&u&M&*&*\\
        4&u&*&*&*\\
    \end{tabular}
    \begin{tabular}{c||ccccc}
        $m\backslash n$&3&4&5&6\\\hline\hline
        2&u&u&u&M\\
        3&u&u&*&*\\
        4&u&M&*&*\\
    \end{tabular}
\end{table}

We identify four minimal configurations: $2r/6s$ and $3r/3s$ in 2D and $2r/6s$ and $4r/4s$ in 3D. It is good to notice that if a network has more receivers \textit{and} transmitters than a minimal configuration, than it can be reduced to it by leaving out the extra nodes, solving the minimal configurations and then solving for the extra nodes.

In the rest of the paper we address the following questions:

\begin{itemize}
    \item Can we solve the minimal configurations algebraically?
    \item How many solutions do the minimal configurations have? How many of these are real?
    \item Can we use the minimal solvers to solve overdetermined networks in noisy environments?
\end{itemize}

\section{Proposed Method}
\label{ch:method}
In this section we first describe the algebraic preprocessing performed on the equations and then briefly review the theory behind \textit{homotopy continuation}, the technique used to solve the polynomial equations.

\subsection{Algebraic preprocessing}

Let $d_{ij}$ denote the distance between the $i$th receiver and $j$th transmitter. Define also
\begin{equation}
    \tilde{d}_{ij}=d_{ij}^2-d_{i1}^2=(f_{ij}-o_j)^2-(f_{i1}-o_1)^2
\end{equation}
for $i=1,\ldots,m$ and $j=2,\ldots,n$. Hence, the term $\tilde{d}_{ij}$ depends on the measurements and offsets. By algebraic manipulation of the previous equation we obtain
\begin{equation}
    -2(\vec{s}_j-\vec{s}_1)\T\vec{r_i}=\tilde{d}_{ij}-\norm{\vec{s}_j}^2+\norm{\vec{s}_1}^2. \label{eq:rec_lin}
\end{equation}
Since we know the transmitter positions, the previous equation is linear in the receivers coordinates. We can use this to eliminate the variables as follows 
\begin{itemize}
    \item \textbf{2D 3r/3s}: we obtain $2$ equations like \eqref{eq:rec_lin} for each receiver. As each receiver has two unknowns, we can solve the receivers as a function of the offsets. Finally, substituting these into the $3$ equations between the $i$th receiver and first transmitter we obtain $3$ equations in $3$ unknown offsets in the form $\norm{\vec{r}_i-\vec{s}_1}^2=(f_{i1}-o_1)^2$. These can be robustly solved by homotopy continuation.
    \item \textbf{3D 4r/4s}: same numerical recipe of 3r/3s
    \item \textbf{2D 2r/4s}: Considering only the first $3$ transmitters, we can repeat the same procedure as that of 3r/3s and obtain $2$ equations in $3$ unknown offsets. Furthermore, noticing that
    \begin{equation}
        \begin{split}
            d_{24}^2-d_{14}^2&=(f_{24}^2-o_4)^2-(f_{14}-o_4)^2\\&=-2(f_{24}-f_{14})o_4+f_{24}^2-f_{14}^2,
        \end{split}
    \end{equation}
    we obtain a linear equation in the $4$th offset and hence also that variable can be eliminated. Adding to the $2$ previously obtained equation the equation $(f_{14}-o_4)^2=\norm{\vec{r}_1-\vec{s}_4}^2$ we finally obtain $3$ equations in $3$ offsets and we can solve the resulting system with homotopy continuation.
    \item \textbf{3D 2r/6s}: same numerical recipe of 2r/4s
\end{itemize}

\subsection{Homotopy Continuation}
Homotopy continuation is a numerical iterative algorithm from algebraic geometry used to solve systems of polynmial equations \cite{morgan1987computing} which has proved itself useful in several applications \cite{malioutov2005homotopy, fabbri2020trplp}.

Let $\vec{F}(\vec{x}):\R^n\rightarrow\R^n$ be a vector of $n$ polynomials in $n$ variables. Our goal is to solve the system $\vec{F}(\vec{x})=\vec{0}$. To do so, we first construct a starting system $\vec{G}(\vec{x})=\vec{0}$ that can be easily solved. The only requirement on $\vec{G}$ is that it must have at least as many distinct solutions as $\vec{F}$. Several techniques to construct such a system exist. In this work, we use the so called \textit{polyhedral initialisation} described in \cite{huber1995polyhedral}. Next, we can define the \textit{homotopy}

\begin{equation}
    \vec{H}(\vec{x}, t) = (1-t)\vec{F}(\vec{x}) + \gamma t\vec{G}(\vec{x}), \label{eq:homotopy}
\end{equation}

where $\gamma$ is a randomly chosen complex number with $\norm{\gamma}=1$ (introduced for numerical stability) and $t$ is a new variable. The key observation is now that the solution of $\vec{H}(\vec{x},t=0)=\vec{0}$ corresponds to the solution of $\vec{F}(\vec{x})=\vec{0}$ and the solution of $\vec{H}(\vec{x},t=1)=\vec{0}$ corresponds to the solution of $\vec{G}(\vec{x})=\vec{0}$. Furthermore, it can be shown that when $t$ varies from $1$ to $0$, the roots of \eqref{eq:homotopy} vary smoothly from the roots of $\vec{G}$ to the roots of $\vec{F}$. This gives a recipe for the homotopy solver: fix a small step $h$ and iteratively solve the equation $\vec{H}(\vec{x},t_k-h)$ using Newton iteration and the solution of $\vec{H}(\vec{x},t_k)$ as initial guess. By the smoothness assumption, the solution at each step will be close to the solution at the previous step and hence the system can be solved efficiently in a few iterations. The homotopy algorithm is initialized with the solution of $\vec{G}(\vec{x})=\vec{0}$ which can be computed efficiently by the  assumptions on $\vec{G}$. In the experiments of this paper, we use the publicly available \textsf{HomotopyContinuation.jl} \cite{timme2018homotopy} library.

\section{Results}
\label{ch:results}
\begin{figure*}[t!]
    \centering
    \begin{subfigure}{0.23\linewidth}
        \includegraphics[width=\linewidth]{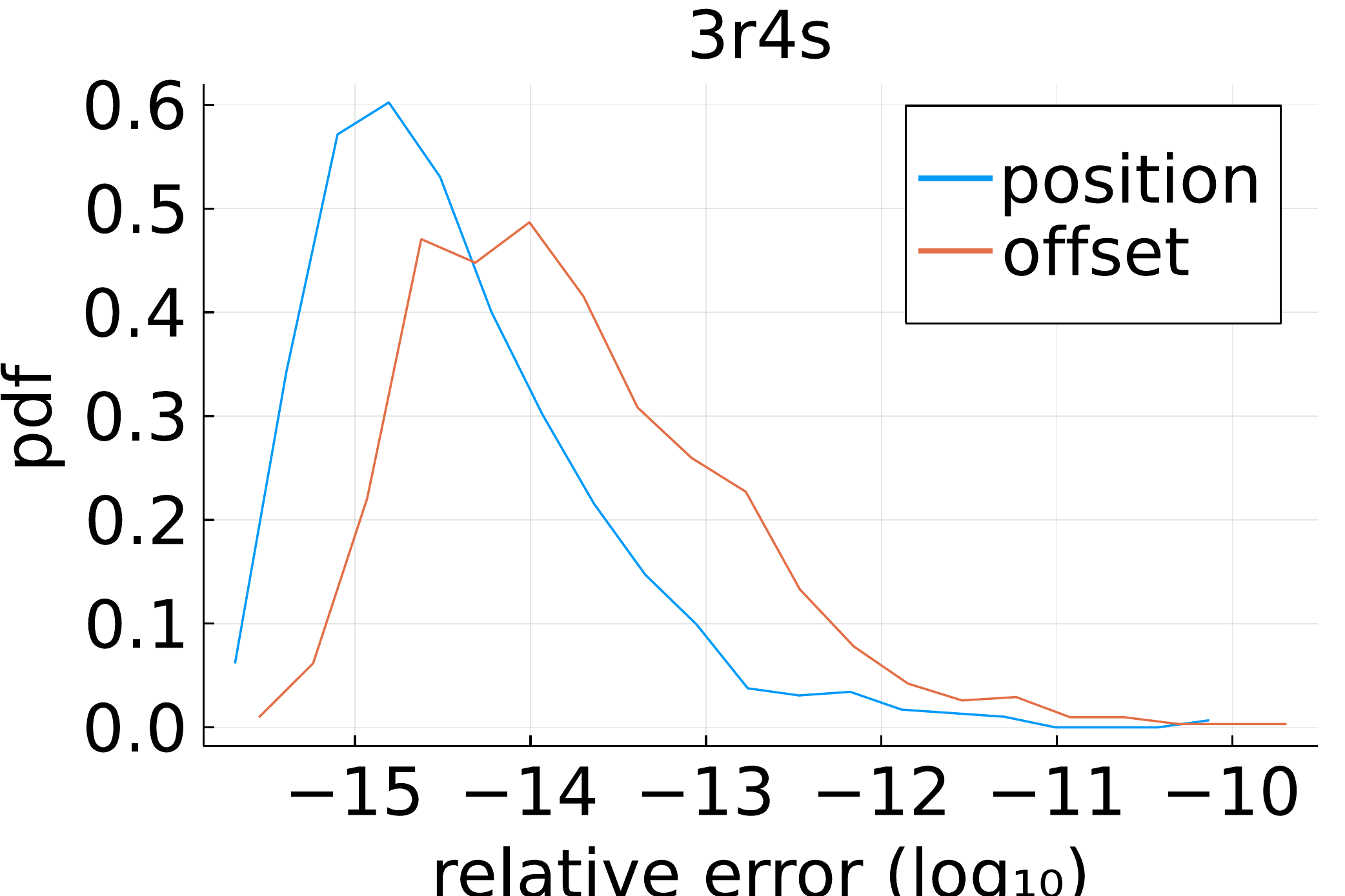}
    \end{subfigure}
    \begin{subfigure}{0.23\textwidth}
        \includegraphics[width=\linewidth]{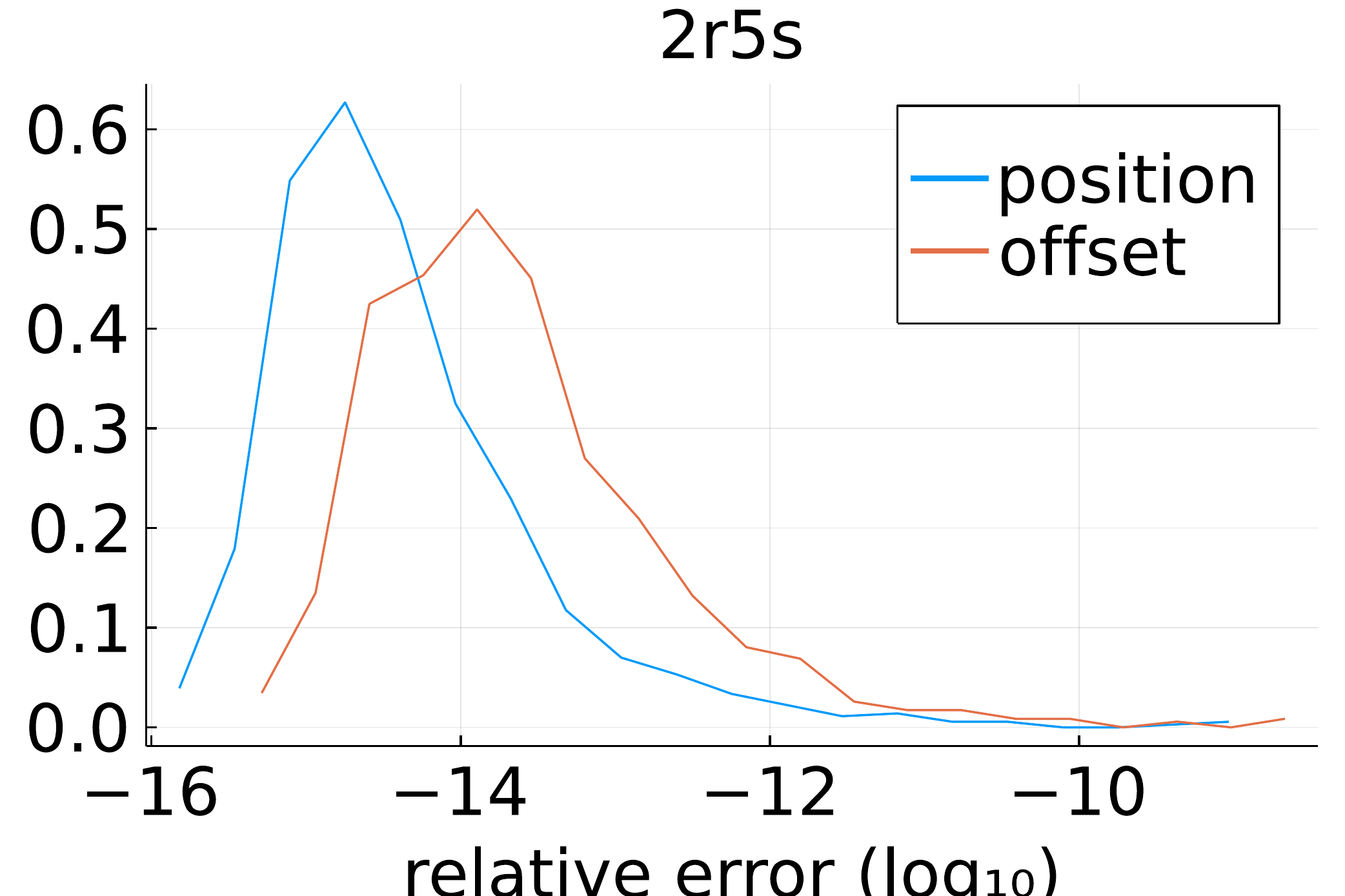}
    \end{subfigure}
    \begin{subfigure}{0.23\textwidth}
        \includegraphics[width=\linewidth]{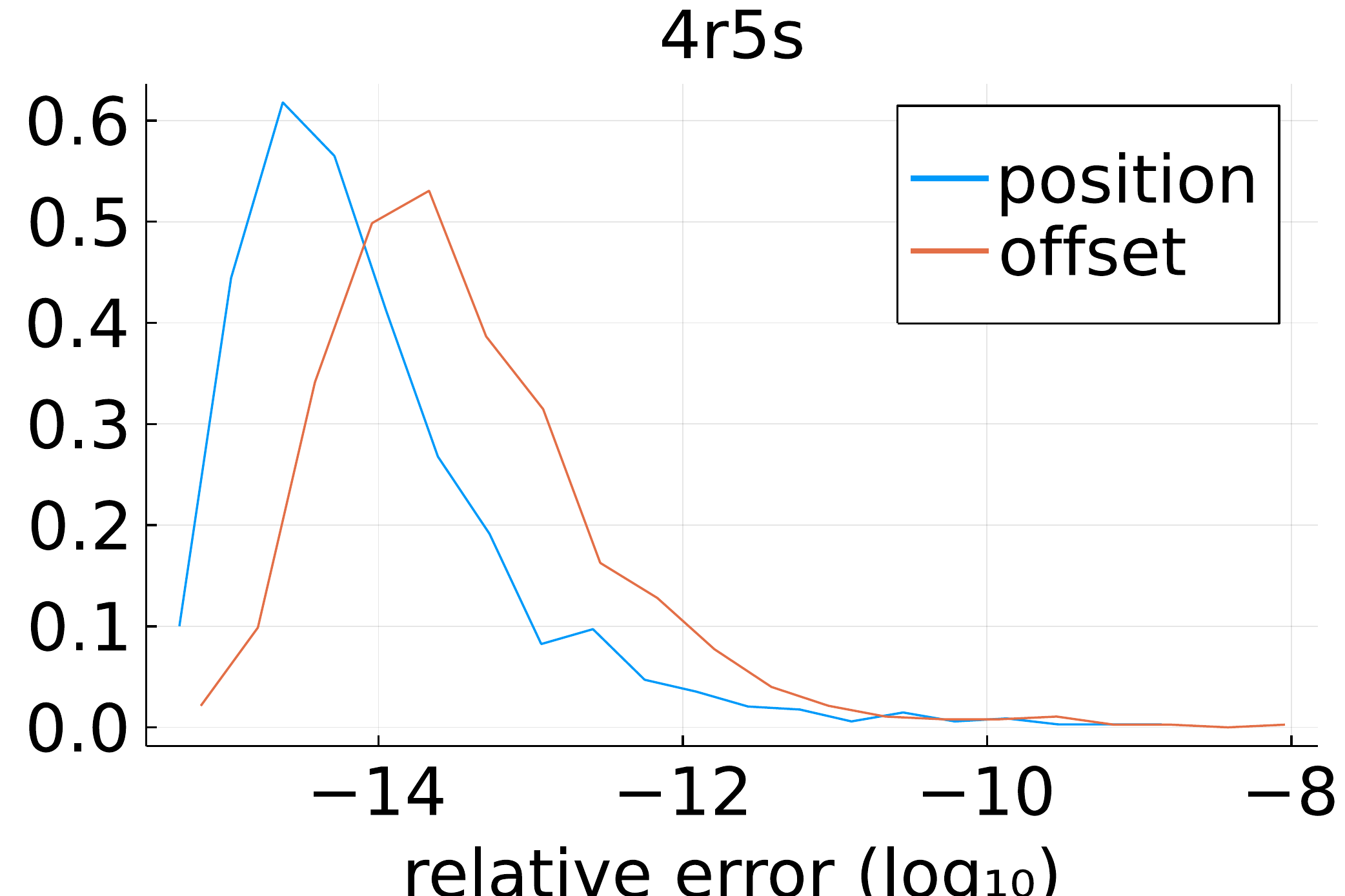}
    \end{subfigure}
    \begin{subfigure}{0.23\textwidth}
        \includegraphics[width=\linewidth]{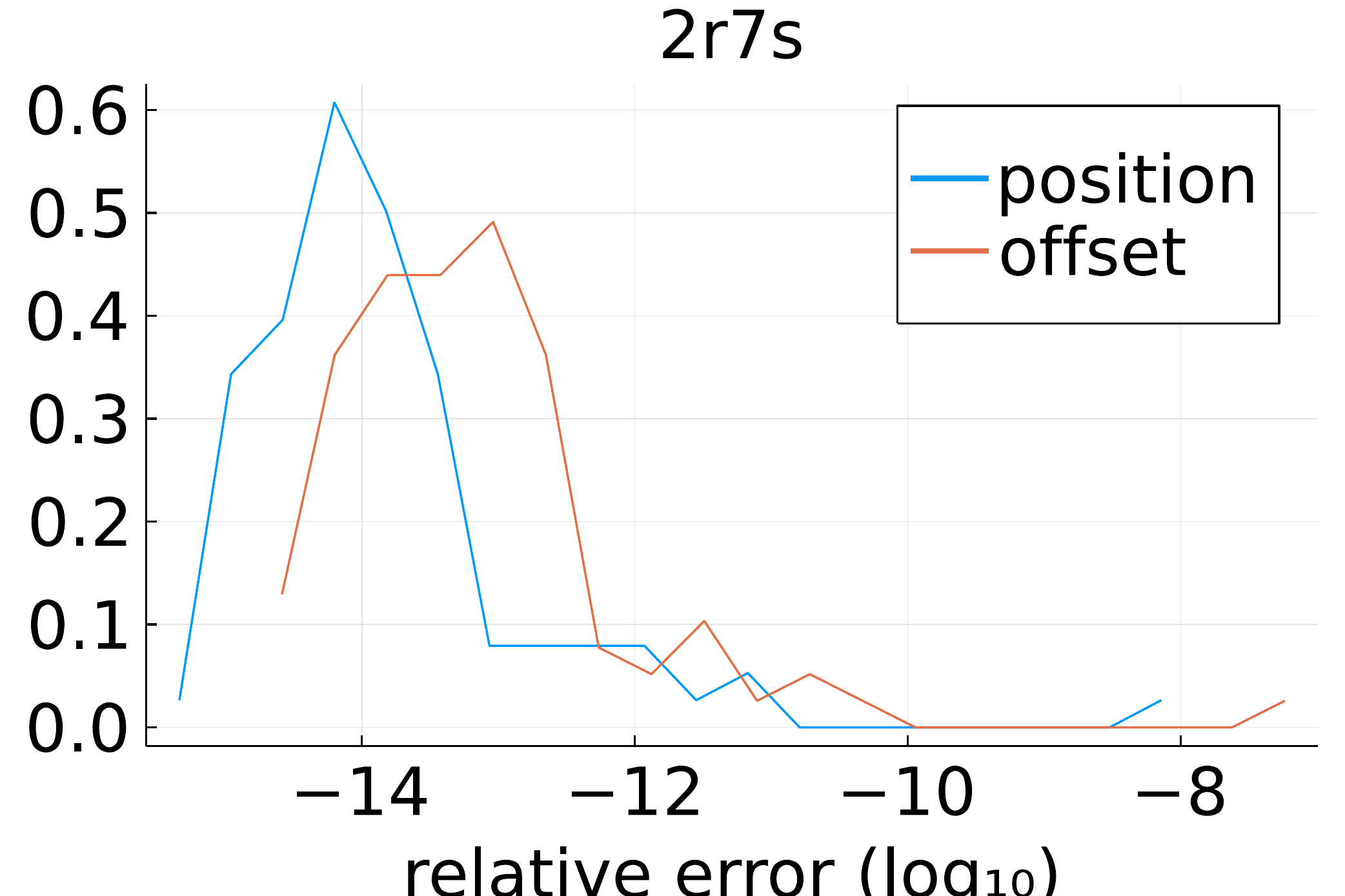}
    \end{subfigure}
    
        \begin{subfigure}{0.23\linewidth}
        \includegraphics[width=\linewidth]{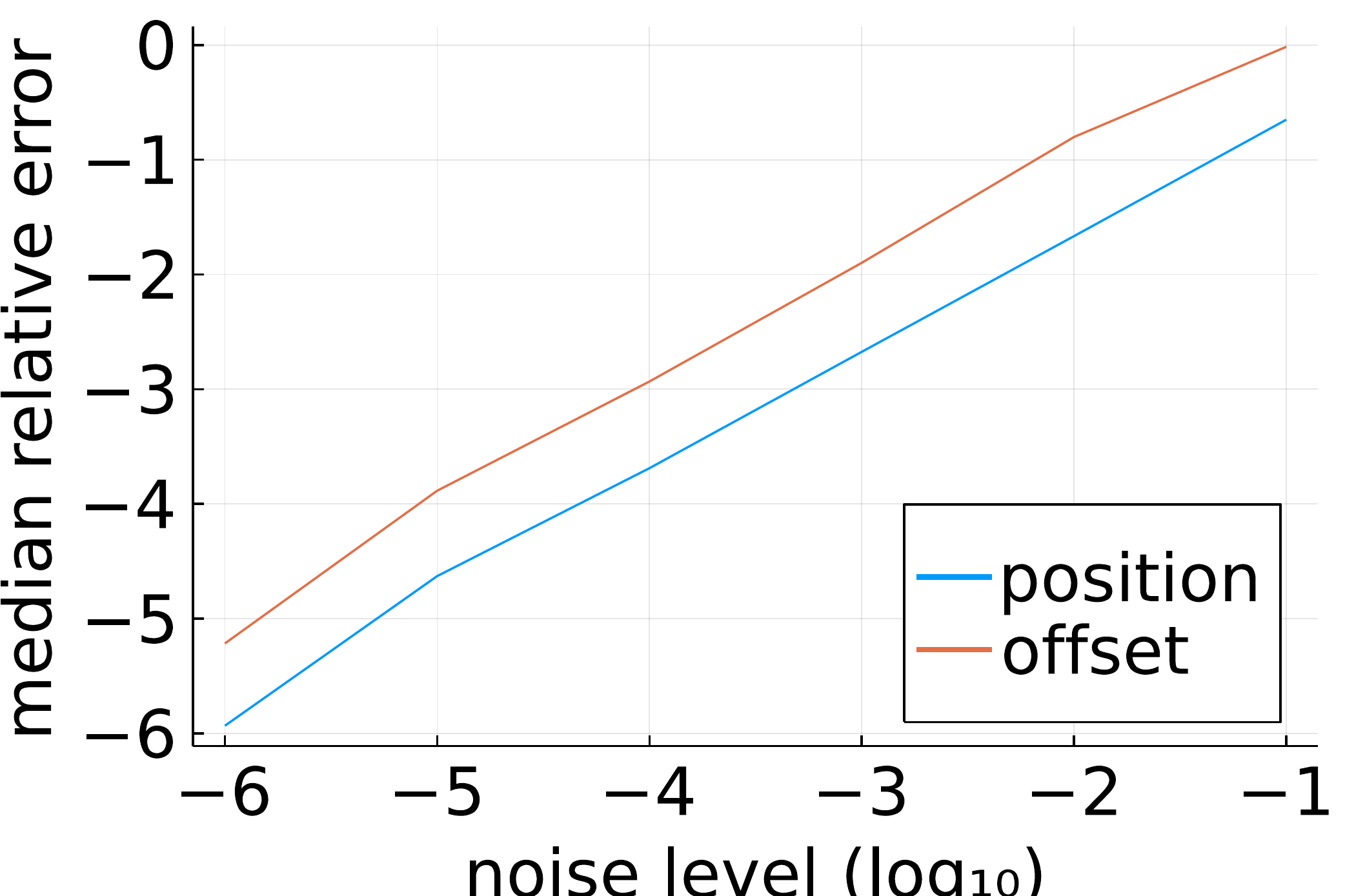}
    \end{subfigure}
    \begin{subfigure}{0.23\textwidth}
        \includegraphics[width=\linewidth]{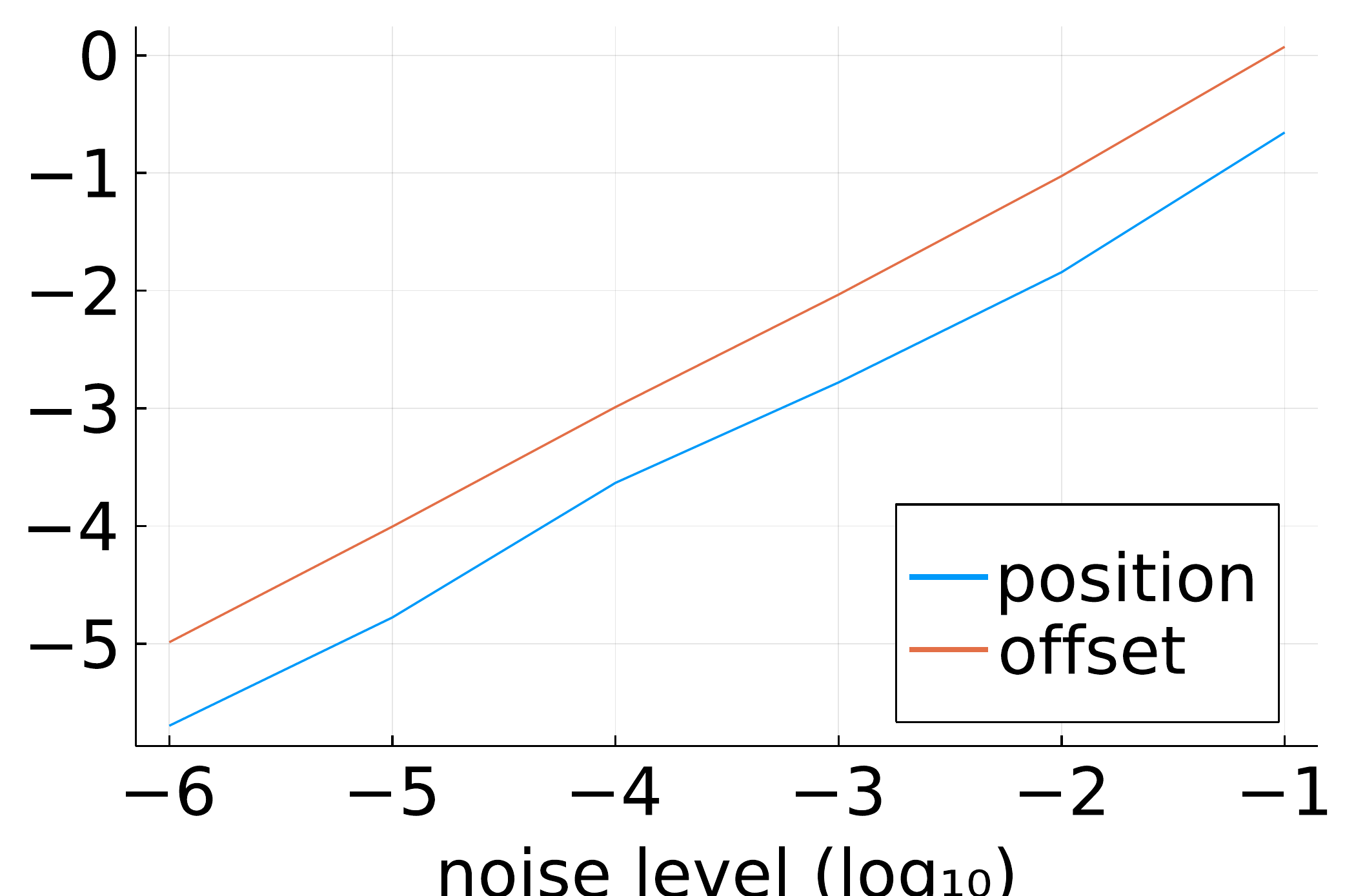}
    \end{subfigure}
    \begin{subfigure}{0.23\textwidth}
        \includegraphics[width=\linewidth]{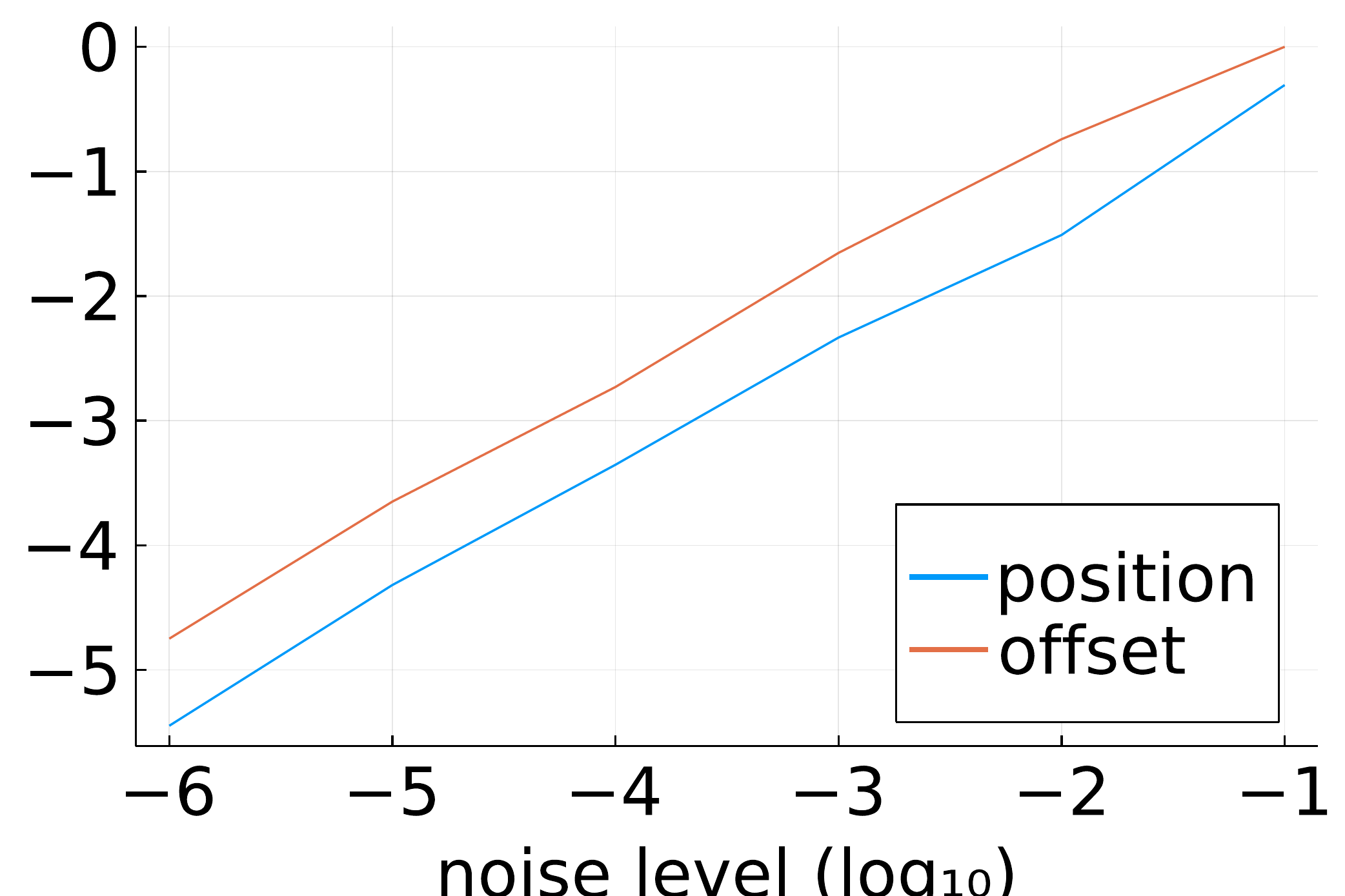}
    \end{subfigure}
    \begin{subfigure}{0.23\textwidth}
        \includegraphics[width=\linewidth]{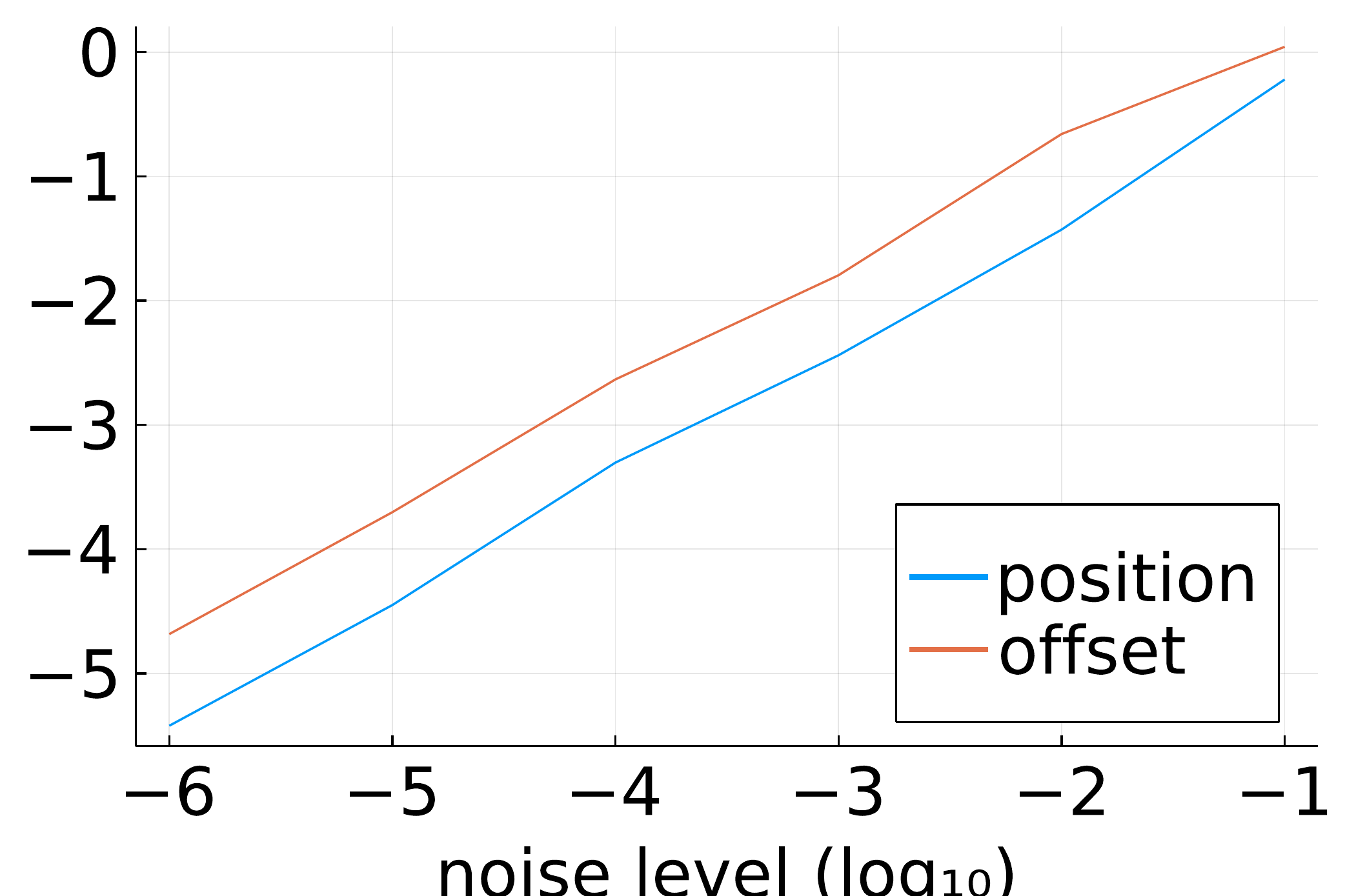}
    \end{subfigure}
    
    \caption{Quantitative benchmark of our solvers. \textbf{Upper row}: relative error distributions for clean data. \textbf{Lower row}: median relative error at different noise levels. Both are in logarithmic scale.}
    \label{fig:results}
\end{figure*}

In this section we present the results of the numerical experiments with the presented solvers. For the experiments conducted with synthetic data, where the coordinates of receivers and transmitters are uniformly sampled from the interval $[-10, 10]$ and offsets are sampled from a standard normal distribution. The experiments are carried on a windows laptop with i7-8565U CPU @ 1.80GHz processor and 8 GB of RAM.

\subsection{Study on the number of solutions}
First, we want to characterise the solutions of the minimal configurations, i.e.\ compute how many (complex) solutions the system of polynomial equations has, and how many of these are real. To determine this, we generated 1000 random instances of the problem and solved with our homotopy solvers. The total number of complex solutions was also computed symbolically using Macaulay2 software \cite{eisenbud2001computations}, so actually proving that the result is correct. The results are reported in Table \ref{tab:num-sol}, where we also report the running time of the homotopy solvers.

\begin{table}[tb]
    \centering
    \caption{Solutions of MOM configurations.}
    \label{tab:num-sol}
    \begin{tabular}{c||c|c|c|c|c}
        &&\multicolumn{3}{c}{Real solutions}\\
         Configuration&tot. sols&min&avg.&max&time [s]\\\hline\hline
         2r/4s 2D&24&4&9&20&0.05\\
         3r/3s 2D&28&2&7&18&0.13\\
         4r/4s 3D&92&2&4&20&0.6\\
         2r/6s 3D&48&4&7&20&3\\
    \end{tabular}

\end{table}
As could be expected by their nonlinearity, MOM configurations don't have a unique solution. It is however interesting to notice that the number of feasible (i.e. real) solutions is strictly smaller than the total number of solutions. Particularly, the 4r/4s presents the highest number of solutions, but the percentage of real solutions is significantly smaller.

\subsection{Solvers to find a unique solution}

Without further information, it is not possible to determine which of the real solutions correspond to the original network configuration. If a unique solution is desired, then at least one extra point needs to be added. Hence we consider now the subminimal configurations with one extra transmitter, that is 3r/4s and 2r/5s in 2D and 4r/5s and 2r/7s in 3D. These problems can be solved as follows: first solve the corresponding minimal configuration by leaving the last transmitter out. Next, for each candidate solution compute the extra offset as the average of the offsets computed with the $m$ equations in form \eqref{eq:range} corresponding to the extra transmitter. Finally, substitute the full solutions in the original equations and choose as final estimate the one with the smallest residual error. 

The proposed solvers were benchmarked with both clean and noisy data. For clean data, we show the error distributions for the relative errors in the histograms in Figure \ref{fig:results}. For noisy data, additive white Gaussian noise with variable variance was added to the measurements before solving. The median relative error as function of the noise level is depicted in the lower row of Figure \ref{fig:results}. As the figure shows, the proposed solvers are stable and robust to noise. 

\subsection{Real data}

We also evaluated our system using real data. The setup consisted of 12 ($m=12$) omni-directional microphones (the T-bone MM-1) spanning a volume of $4.0 \times 4.6 \times 1.5$ meters.
%(see \figurename~\ref{fig:setup}).
A speaker was moved through the setup while emitting sound.
Ground truth positions for the microphones and speaker positions were found using a Qualisys motion capture system.
The microphones were all internally synchronized, but we assume that the time of sound emission from the speaker is unknown. We use the position estimation of the sound sources from the Qualisys system. Consequently, the microphone positions and emission times correspond to the situation of unknown receivers and offsets, while the sender positions are assumed to be known. In the experiment a song was played through the speaker and the arrival times $t_{ij}$ were found using GCC-PHAT \cite{Knapp1976}. This resulted in a total of $n=151$ sound events with available pseudoranges. Next, we sampled $4$ receivers and $5$ transmitters and solved the problem using our $4r/5s$ solver. Finally, we solved the remaining offsets as described above and trilaterated the remaining receivers. This estimate was further finalized using Levenberg-Marquardt algorithm. As a final result, the mean position error for the receivers was $\SI{10}{\centi\meter}$.

\section{Conclusions}
\label{ch:conclusions}
In this paper we proposed a new framework, Multiple Offsets Multilateration, to compute receivers positions from radio measurements from reference transmitters at known positions, which are however unsynchronized. We derived a mathematical formulation of this new framework and presented a full characterization both in 2D and 3D, identifying what are the minimal configurations, i.e.\ how many nodes the network must have to be solvable. The numerical experiments we presented have both theoretical and practical importance. From the theoretical side, we determined symbolically using algebraic geometry the total number of complex solutions each minimal configuration can have. We also gave empirical results of the number of real solutions. This is important to understand the computational complexity of the problem. As a more practical contribution, we used homotopy continuation to derive efficient and robust polynomial solvers for the minimal configurations and showed how these can be used to obtain accurate estimates also in noisy environments.

\subsection*{Acknowledgment}
This work was partially funded by the Academy of Finland project 327912 REPEAT and the Swedish strategic research project ELLIIT.The authors gratefully acknowledge Lund University Humanities Lab.

\bibliographystyle{IEEEbib}
\bibliography{ref}

\end{document}